\documentclass[epj,nopacs]{svjour}
\usepackage[english]{babel}
\usepackage[T1]{fontenc}
\usepackage{bm,bbm}
\usepackage{amssymb}
\usepackage{amsmath}
\usepackage{graphicx} 
\usepackage{subfigure}
\usepackage{multirow}
\usepackage{isotope}
\usepackage{cite}

\begin{document}

\title{Medium-mass hypernuclei and the nucleon-isospin dependence of the three-body hyperon-nucleon-nucleon force}

\author{Diego Lonardoni\inst{1,2} \and Francesco Pederiva\inst{3,4}}

\authorrunning{D. Lonardoni and F. Pederiva}
\titlerunning{Medium-mass hypernuclei and the isospin dependence of the three-body $\Lambda NN$ force}

\institute{National Superconducting Cyclotron Laboratory, Michigan State University, East Lansing, Michigan 48824 \and Theoretical Division, Los Alamos National Laboratory, Los Alamos, New Mexico 87545 \and Physics Department, University of Trento, via Sommarive 14, I-38123 Trento, Italy \and INFN-TIFPA, Trento Institute for Fundamental Physics and Application, Trento, Italy}

\date{Received: XXX / Revised version: XXX}

\abstract{We report quantum Monte Carlo calculations of single-$\Lambda$~hypernuclei for $A<50$ based on phenomenological two- and three-body hyperon-nucleon forces. We present results for the $\Lambda$~separation energy in different hyperon orbits, showing that the accuracy of theoretical predictions exceeds that of currently available experimental data, especially for medium-mass hypernuclei. We show the results of a sensitivity study that indicates the possibility to investigate the nucleon-isospin dependence of the three-body hyperon-nucleon-nucleon force in the medium-mass region of the hypernuclear chart, where new spectroscopy studies are currently planned. The importance of such a dependence for the description of the physics of hypernuclei, and the consequences for the prediction of neutron star properties are discussed.} 

\maketitle

\section{Introduction}
The recent observation of gravitational waves from the merger of two neutron stars (NSs)~\cite{AbbottPRL:2017,AbbottAPJ:2017} provides new constraints on the equation of state (EoS) of high density matter. An important example is the tidal deformability parameter, which is connected to the EoS of matter in the NS core~\cite{Fattoyev:2018}. This matter is characterized by a density several times larger than the typical density of a nucleus, and by an extreme nucleon asymmetry, i.e., the number of neutrons largely exceeds that of protons. 

While it would be desirable to have a theory of NS structure directly based on quantum chromodynamics, at present we still need to rely on effective theories using nucleons (and possibly mesons) as active degrees of freedom~\cite{Wiringa:1995,Epelbaum:2009,Machleidt:2011,Ekstrom:2013,Gezerlis:2014,Entem:2015,Epelbaum:2015,Ekstrom:2015,Piarulli:2015,Lynn:2016,Ekstrom:2017}. These have proven to be remarkably successful in the description of low-energy properties of nuclei~\cite{Barrett:2013,Hagen:2013,Carlson:2015,Hergert:2015,Lonardoni:2018prc}, and they have also been employed for the prediction of NS structure~\cite{Akmal:1998,Hebeler:2013,Carlson:2015}.

In the inner core of a NS, Pauli blocking might favor the appearance of heavier baryons, such as hyperons $(Y)$~\cite{Ambartsumyan:1960}. Hyperons are expected to make the EoS softer, reducing the maximum mass that a NS can stably support. However, different models predict very different results $(1.3M_\odot\lesssim M_{\max}\lesssim 2M_\odot)$~\cite{Vidana:2011,Schulze:2011,Massot:2012,Bednarek:2012,Miyatsu:2013,Lopes:2014,Yamamoto:2014,Lonardoni:2015,Fortin:2015}, and there is no clear consistency with the observation of heavy NSs~\cite{Demorest:2010,Antoniadis:2013}.

This puzzle can only be solved by performing accurate many-body calculations based on a well constrained theory of the hyperon-nucleon $(YN$) and hyperon-hyperon $(YY)$ interactions. Unfortunately, both $YN$ and $YY$ interactions are not well constrained by current experiments. Direct $YN$ scattering is technically difficult to perform, and few data are available~\cite{Gal:2016}. Some information is available for $YY$ scattering, like the derivation of the $\Lambda\Lambda$ scattering length from heavy-ion collisions~\cite{Morita:2015}, and the Nagara event~\cite{Takahashi:2001}, but it only allows for constraining at most a central contact interaction. Moreover, as in the non-strange sector, many-body hypernuclear forces ($YNN$, $YYN$, \ldots) are expected to be relevant, and need to be constrained. Experimental information on hypernuclei---bound states of nucleons and hyperons---is thus the only basis for the construction of a realistic hyperon-nucleon potential. Several attempts at using a theory that includes the whole baryon octet, and accounts for the so called $\Lambda N$-$\Sigma N$ coupling, have been made~\cite{Rijken:2010,Rijken:2013,Haidenbauer:2013,Haidenbauer:2017,Petschauer:2017}. However, these schemes need a relatively large number of parameters that are presently difficult to access experimentally  and require extra theoretical constraints.

In Refs.~\cite{Lonardoni:2013,Lonardoni:2014}, a phenomenological model for the $\Lambda N$ and $\Lambda NN$ forces~\cite{Bodmer:1984,Usmani:1995_3B}, inspired by the phenomenological $NN$~\cite{Wiringa:1995,Wiringa:2002} and $NNN$~\cite{Pudliner:1997,Pieper:2001,Pieper:2008} potentials, was derived using quantum Monte Carlo algorithms. This interaction was introduced with the precise aim of building a realistic model that i) is tightly connected to the experimentally available information and ii) contains the least number of parameters possible. Its success was shown by the capability of accurately describing the properties of $\Lambda$~hypernuclei in a wide mass range~\cite{Lonardoni:2013,Lonardoni:2014}. In order to make a first connection to NS properties, the same interaction was successively employed to study hyper-neutron matter~\cite{Lonardoni:2015}. 

\begin{figure*}[t]
\centering
\subfigure[\label{fig:ln}]{\includegraphics[height=3.5cm]{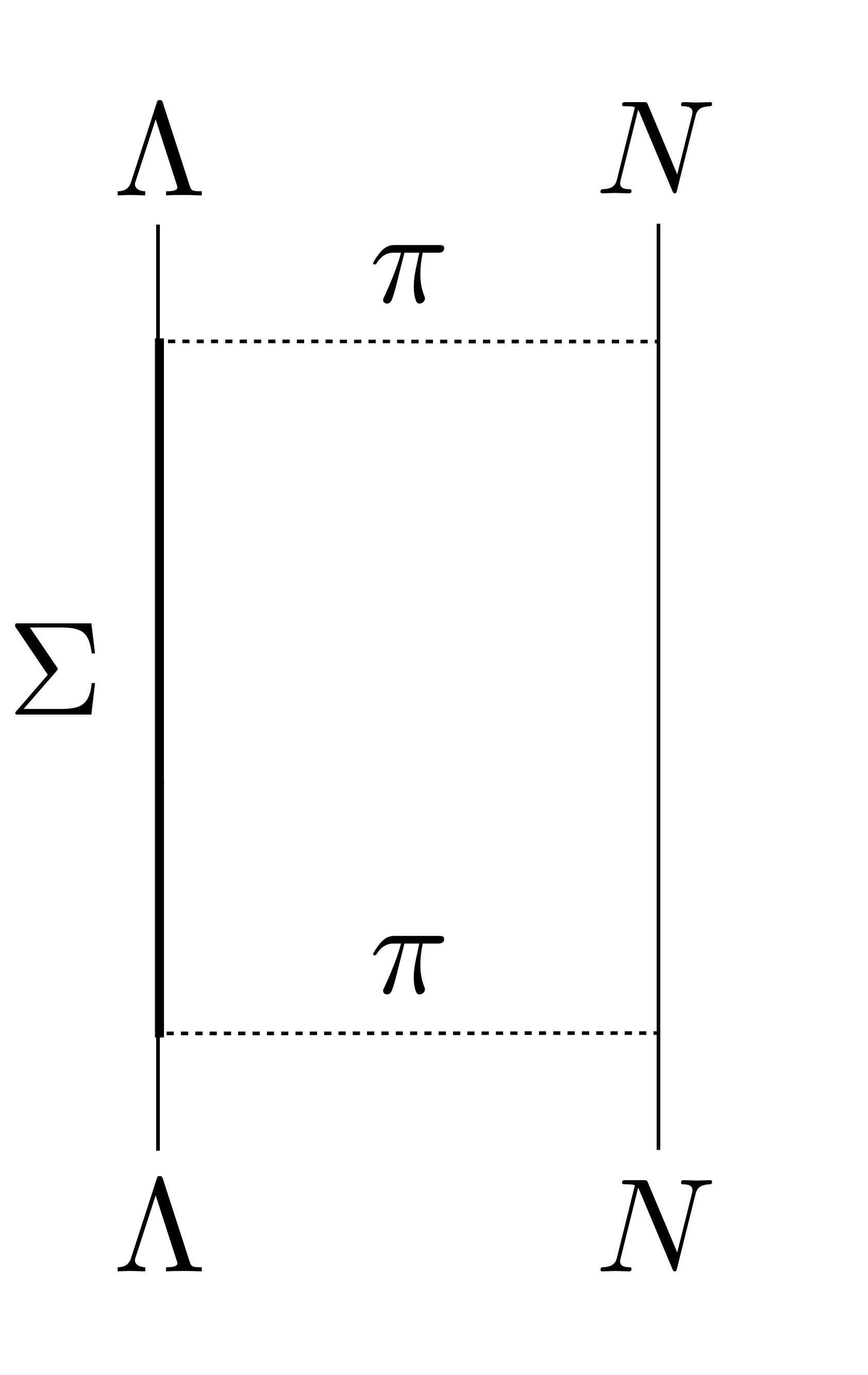}}\qquad\qquad\quad
\subfigure[\label{fig:lnn_p}]{\includegraphics[height=3.5cm]{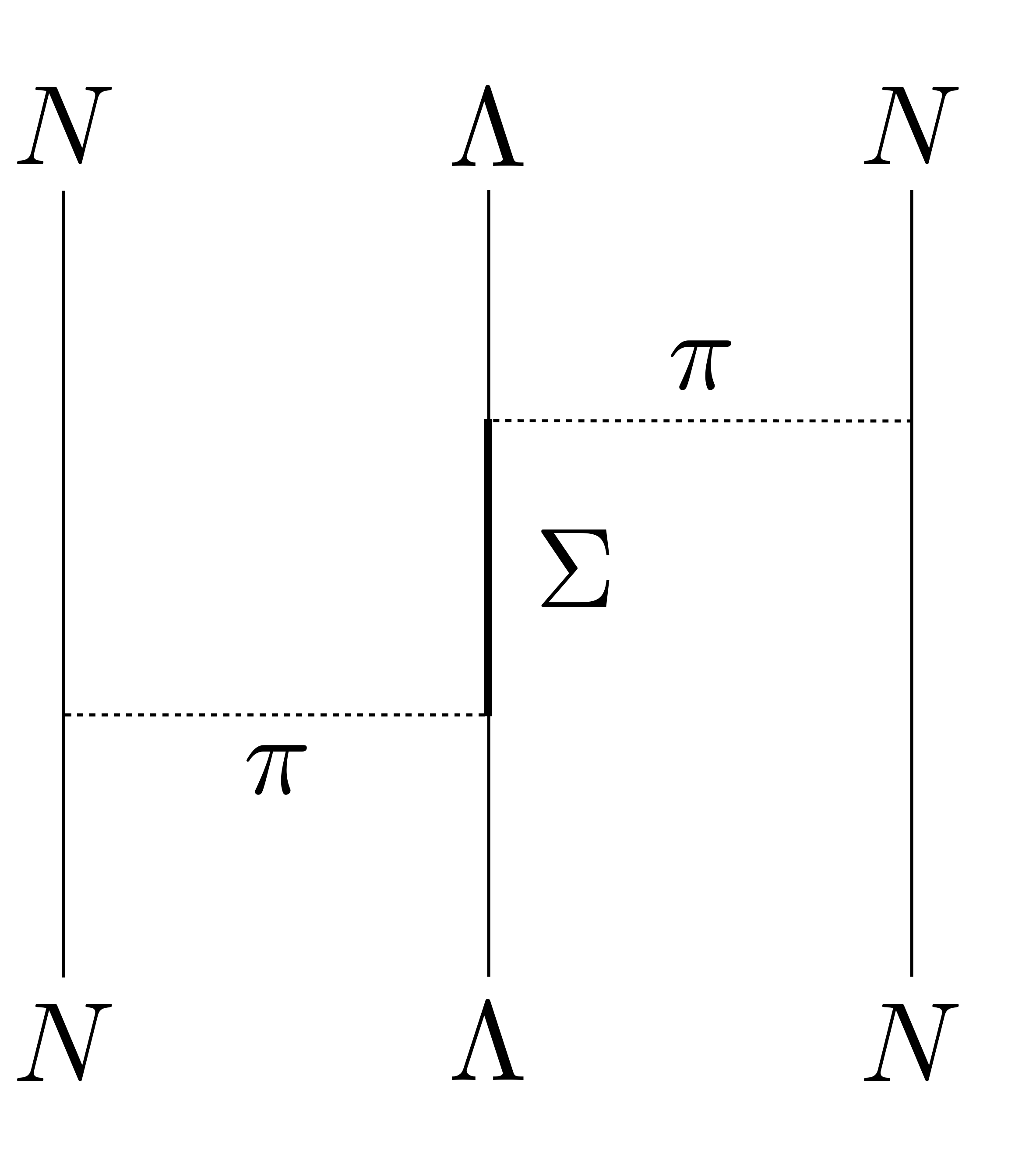}}\qquad\qquad\quad
\subfigure[\label{fig:lnn_s}]{\includegraphics[height=3.5cm]{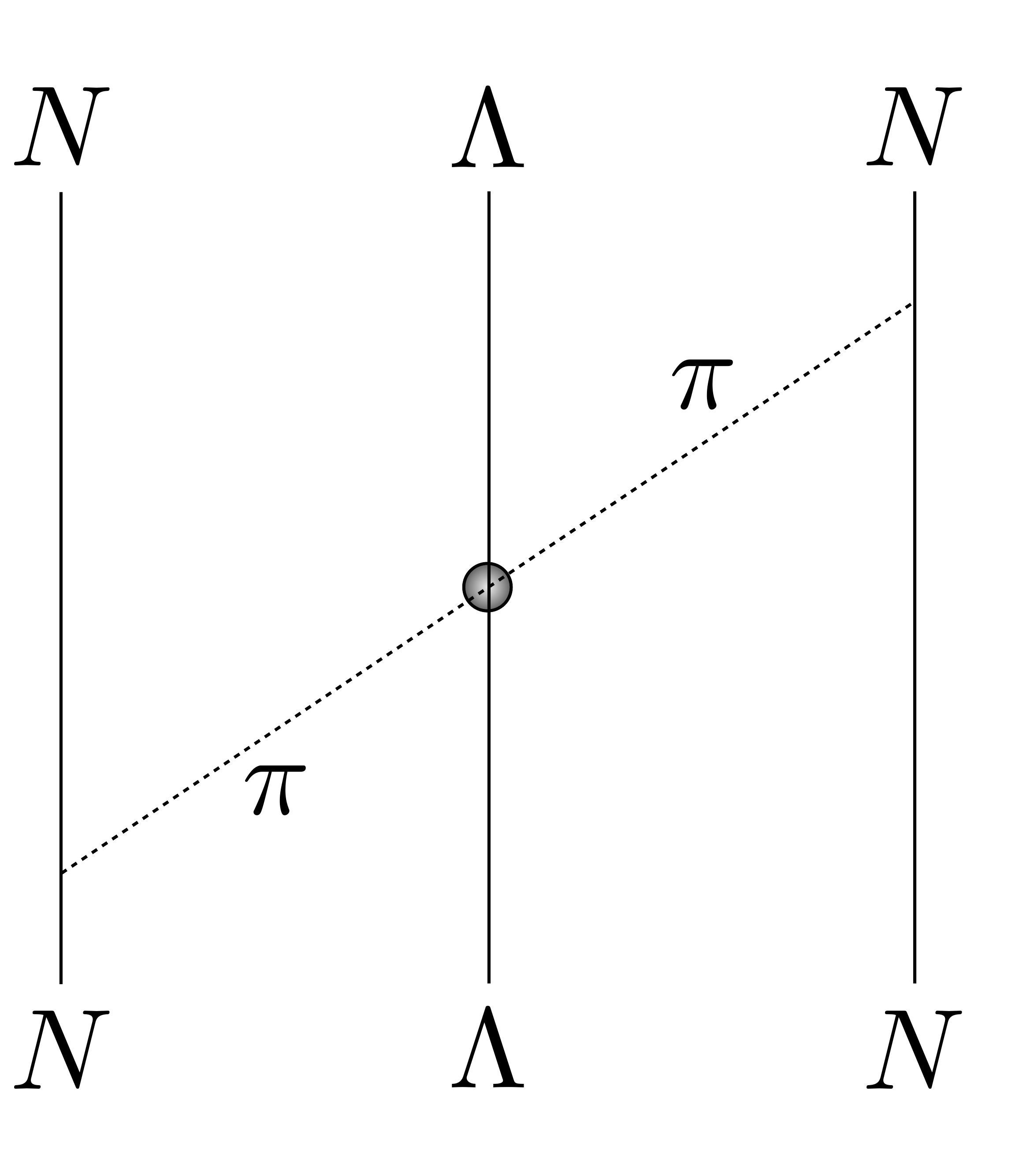}}\qquad\qquad\quad
\subfigure[\label{fig:lnn_d}]{\includegraphics[height=3.5cm]{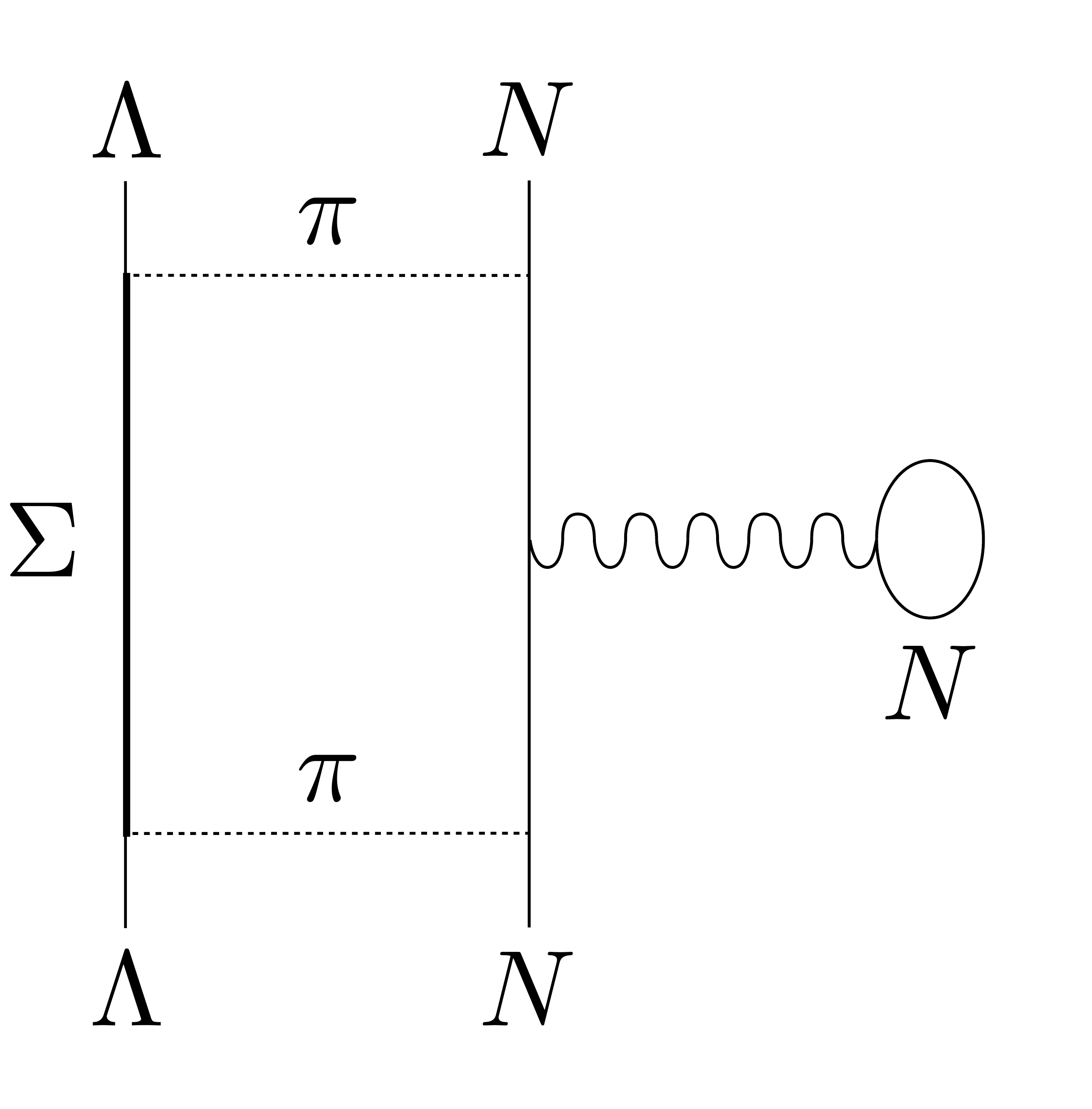}}
\caption[]{Hyperon-nucleon two-pion exchange diagrams. 
	Panel (a) represents the two-body $\Lambda N$ channel.
	Panels (b)-(d) are the three-body $\Lambda NN$ channels.}
\label{fig:diagrams}
\end{figure*}

In this paper, we address an additional crucial point. In a hypernucleus, contributions to the binding energy from $YNN$ interactions, as they are introduced in our scheme, might be dominated by contributions of $\Lambda pn$ triplets, mainly due to the Pauli principle suppressing $\Lambda nn$ ($\Lambda pp$) contributions. However, $\Lambda nn$ terms are expected to be largely dominant in the NS core. By means of the auxiliary field diffusion Monte Carlo (AFDMC) method~\cite{Schmidt:1999}, we try to assess the extent of the nucleon-isospin dependence of these three-body contributions by studying the ground state of medium-mass hypernuclei. In particular, we provide predictions for $A=40,48$ potassium hypernuclei, the study of which is the goal of the current hypernuclear experimental program at the Thomas Jefferson National Accelerator Facility (JLab)---approved experiment E12-15-008~\cite{JLab:2016}. This is an important step toward a coherent description of hypernuclear experiments and the phenomenology related to NSs, eventually including the information obtained from gravitational waves observations.

\section{Hamiltonian}
The Hamiltonian for single-$\Lambda$~hypernuclei \isotope[A][\Lambda]{Z}, with $A$ the total number of baryons, is of the form:
\begin{align}
	H=	& -\frac{\hbar^2}{2m_N}\sum_i \nabla_i^2+\sum_{i<j}v_{ij}+\sum_{i<j<k}v_{ijk}+ \nonumber \\ 
    	& -\frac{\hbar^2}{2m_\Lambda}\nabla_\lambda^2+\sum_{i}v_{\lambda i} +\sum_{i<j}v_{\lambda ij},
	\label{eq:ham} 
\end{align}
where Latin indexes $\{i,j,k\}$ label nucleons, and $\lambda$ is used for the $\Lambda$ particle. The first line of Eq.~(\ref{eq:ham}) corresponds to the Hamiltonian of the core nucleus \isotope[A-1]{Z}, and the second line contains the kinetic energy of the $\Lambda$ particle and two- and three-body interactions among hyperon and nucleons. 

Realistic two- and many-body forces are in general necessary for an accurate description of the binding energies of nuclei and hypernuclei. Although the $\Lambda$~separation energy, defined as the energy difference between the core nucleus and the hypernucleus,
\begin{align}
	B_\Lambda\left(\isotope[A][\Lambda]{Z}\right)=E\left(\isotope[A-1]{Z}\right)-E\left(\isotope[A][\Lambda]{Z}\right)\,, 
\end{align}
is greatly affected by the employed hypernuclear potential, it is mostly insensitive to the details of the nuclear interaction~\cite{Lonardoni:2013}. For this reason we consider here somewhat simplified two- plus three-nucleon potentials, supported by realistic two- and three-body hyperon-nucleon forces. The advantage of this choice is a much simpler implementation in our computational method that allows us to perform accurate calculations of the $\Lambda$~separation energy of medium-mass systems at an acceptable computational cost.

In the non-strange sector we consider the phenomenological two-nucleon potential Argonne $v_4'$ (AV4')~\cite{Wiringa:2002}, supported by the purely central component of the three-nucleon Urbana IX (UIX) force~\cite{Pudliner:1997}. No Coulomb interaction is considered at this time. AV4' is a simplified version of the more sophisticated Argonne $v_{18}$ (AV18) potential~\cite{Wiringa:1995}, obtained by re-projecting the full interaction onto the first four channels in order to preserve the phase shifts of lower partial waves and the properties of deuterium. It typically overbinds light nuclei~\cite{Wiringa:2002}. We use here the central component of UIX as a source of repulsion to compensate for such an excessive attraction. As we will discuss in the results section, we verified that this simple interaction provides binding energies that are well within 10\% from the experiment.

In the hyperon-nucleon channel, we consider the phenomenological $\Lambda$-nucleon potential introduced in the '80s by Bodmer et al.~\cite{Bodmer:1984}, and employed thereafter in variational ~\cite{Usmani:1995_3B,Usmani:1995,Shoeb:1998,Usmani:1999,Shoeb:1999,Usmani:2008,Imran:2014} and diffusion Monte Carlo calculations~\cite{Lonardoni:2013,Lonardoni:2014,Lonardoni:2015}. Such interaction is expressed in coordinate space and it includes two- and three-body hyperon-nucleon components with a hard-core repulsion between baryons and a spin/isospin operator structure that resembles that of phenomenological two- and three-nucleon potentials~\cite{Wiringa:1995,Pudliner:1997}. We only consider nucleons and $\Lambda$~particles as active degrees of freedom. Since the $\Lambda$~hyperon has zero isospin, no one-pion exchange contributions are allowed due to isospin conservation. As shown in Fig.~\ref{fig:diagrams}, the relevant diagrams including a hyperon and one or two nucleons involve two-pion exchange, with the (possible) excitation of virtual $\Sigma$ state. Differently from other models of the hypernuclear force, where the full baryon octet is involved and the $\Lambda N$-$\Sigma N$ coupling is explicitly introduced~\cite{Rijken:2010,Rijken:2013,Haidenbauer:2013,Haidenbauer:2017,Petschauer:2017}, in our scheme such a coupling is taken into account in an effective way only, by fitting the strength of the contributions coming from the diagrams of Fig.~\ref{fig:diagrams} to available $\Lambda$ hypernuclear experimental data. We do not explicitly consider here effects due to kaon exchange, which would in turn imply the inclusion of terms describing hyperon-nucleon exchange.

The two-body component $v_{\lambda i}$ has been fit to the available $\Lambda p$ scattering data at low energy~\cite{Dalitz:1972,Schimert:1980,Bodmer:1984}, 
and it includes a central and a hyperon-nucleon spin component~\cite{Lonardoni:2014}. A simple two-body spin-independent charge symmetry breaking term (CSB), originally fit to reproduce the energy difference of the $A=4$ mirror hypernuclei from emulsion data, is also included. However, its effect is only appreciable for light systems~\cite{Lonardoni:2014}. Recent experimental analysis~\cite{Esser:2015,Yamamoto:2015,Botta:2017} showed inconsistencies with the old emulsion data, which exhibits a strong spin-dependence of the CSB in $\Lambda N$ interaction. This could be interpreted as a consequence of the $\Lambda$-$\Sigma^0$ mixing mechanism~\cite{Gazda:2016PRL,Gazda:2016}, which could be effectively translated into a three-body $\Lambda NN$ force with explicit nucleon-isospin dependence. This possibility will be investigated in a future study of $s$- and $p$-shell hypernuclei. 

The three-body force $v_{\lambda ij}$ reads
\begin{align}
	v_{\lambda ij}=	&-\frac{C_P}{6}\Big\{X_{\lambda i},X_{\lambda j}\Big\}\,\bm\tau_i\cdot\bm\tau_j \nonumber \\
    				&+C_S\,Z_{\lambda i}\,Z_{\lambda j}\,\bm\sigma_i\cdot \hat{{\bf r}}_{\lambda i}\,\bm\sigma_j\cdot \hat{{\bf r}}_{\lambda j}\,\bm\tau_i\cdot\bm\tau_j \nonumber \\
                    &+W_D\,T_{\lambda i}^2\,T_{\lambda j}^2\left[1+\frac{1}{6}\bm\sigma_\lambda\cdot(\bm\sigma_i+\bm\sigma_j)\right],
	\label{eq:vlij}
\end{align}
where the functions $X_{\lambda i}$, $Z_{\lambda i}$, and $T_{\lambda i}$ are defined in Eqs.~[14], [15], and [8] of Ref.~\cite{Lonardoni:2014}, respectively. The coefficients $C_P$, $C_S$, and $W_D$ have been fit to simultaneously reproduce the $\Lambda$~separation energy of the closed-shell hypernuclei \isotope[5][\Lambda]{He} and \isotope[17][\Lambda]{O}~\cite{Lonardoni:2014}. In this work, we employ a slightly different parametrization $(C_P=1.30\,\text{MeV},C_S=1.50\,\text{MeV},W_D=0.033\,\text{MeV})$ tuned to reproduce the updated $\Lambda$~separation energy of medium-mass hypernuclei. In fact, all the $(\pi^+,K^+)$ data suffer normalization issues, and need to be shifted by $\approx0.5-0.6\,\rm MeV$ in order to be consistent with other production mechanisms (see Refs.~\cite{Gogami:2016,Gal:2016} for a detailed discussion). The aforementioned parameters of the three-body $\Lambda NN$ force have been adjusted to fit the $s$-orbit $B_\Lambda$ of \isotope[5][\Lambda]{He} from emulsion data, and the updated $(\pi^+,K^+)$ values for \isotope[16][\Lambda]{O} and \isotope[40][\Lambda]{Ca}.

\section{AFDMC method}
We compute binding energies of nuclei and hypernuclei by using the auxiliary field diffusion Monte Carlo method. A complete description of the AFDMC algorithm for standard nuclear systems is given in~\cite{Schmidt:1999,Carlson:2015,Lonardoni:2018prc}. The extension of the AFDMC method for strange nuclear systems has been discussed in Refs.~\cite{Lonardoni:2013,Lonardoni:2014,Lonardoni:2015}. In the following we summarize the key features of the algorithm, providing details on the improvements achieved for the computation of the ground state of medium-mass systems.

In the AFDMC method, the many-body Schr\"odinger equation is solved by enhancing the ground-state component of a starting trial wave function using the imaginary-time projection method. The trial state for single-$\Lambda$ hypernuclei takes the form:
\begin{align}
	|\Psi\rangle_{J^\pi,T}=\Bigg(\prod_{i<j}f_{ij}\Bigg)\Bigg(\prod_{i}f_{\lambda i}\Bigg)|\Phi\rangle_{J^\pi,T},
    \label{eq:wf}
\end{align}
where $f_{ij}$ and $f_{\lambda i}$ are nucleon-nucleon and hyperon-nucleon spin/isospin-independent correlation functions, respectively. These are derived by retaining the central component of the solution of the Schr\"odinger-like equations in different spin and isospin channels~\cite{Pandharipande:1979} employing the $NN$ or $YN$ two-body potentials:
\begin{align}
	\left[-\frac{\hbar^2}{2\mu}\nabla^2+v_{S,T}(r)+\lambda_{S,T}(r)\right]f_{S,T}(r)=0.
    \label{eq:sch}
\end{align}
This approach was found to generate more accurate spin/isospin-independent correlation functions compared to the direct solution of Eq.~(\ref{eq:sch}) for the  central channel only of the two-body interaction, as done in Ref.~\cite{Lonardoni:2014}. $|\Psi\rangle_{J^\pi,T}$ consists of a sum of Slater determinants of nucleon single-particle states $\phi_\alpha^N$ coupled with the hyperon single-particle state $\phi_\beta^\Lambda$:
\begin{align}
	|\Phi\rangle_{J^\pi,T} = & \Bigg\{\Big[\sum\mathcal C_{J_NM_N}\mathcal D\left\{\phi_\alpha^N({\bf r}_i,s_i)\right\}\Big]_{J_N,T_N} \nonumber \\
& \times\Big[\phi_\beta^\Lambda({\bf r}_\lambda,s_\lambda)\Big]_{J_\Lambda,0}\Bigg\}_{J^\pi,T},
\end{align}
where ${\bf r}_i$ (${\bf r}_\lambda$) are the spatial coordinates of nucleons (hyperon), and $s_i$ ($s_\lambda$) represents their spinors. $J$ is the angular momentum, $M$ its projection, $T$ the isospin, and $\pi$ the parity. Subscripts in the quantum numbers differentiate nucleons from the $\Lambda$~particle. The nucleon determinants $\mathcal D$ are coupled with Clebsch-Gordan coefficients $\mathcal C_{J_NM_N}$ in order to reproduce the proper quantum numbers of the core nucleus \isotope[A-1]{Z}. These are then coupled with the single-particle sate of the hyperon in order to reproduce the experimental (or expected) total angular momentum, isospin, and parity $(J^\pi,T)$ of the hypernucleus \isotope[A][\Lambda]{Z}. 

Each single-particle state consists of a radial function solution of the self-consistent Hartree-Fock problem with the Skyrme effective interactions of Ref.~\cite{Bai:1997}, coupled to the spin/isospin state. For the $\Lambda$ particle we use the neutron radial functions coupled to the spin state only. The proper hyperon orbital is chosen to construct hypernuclei with the $\Lambda$~particle in the $s$-, $p$-, and $d$-orbit. The free parameters involved in the solution of Eq.~(\ref{eq:sch}) (see Ref.~\cite{Pandharipande:1979} for details), together with scaling factors multiplying the radial functions, are chosen in order to minimize the variational expectation value of the Hamiltonian of Eq.~(\ref{eq:ham}). 

The AFDMC imaginary time propagation for strange systems involves the diffusion in real space of both nucleon and hyperon spatial coordinates. Spinors are also sampled via the Hubbard-Stratonovich transformation, that is applied to the imaginary time propagator for potentials that are quadratic in the spin/isospin operators. The three-body coordinate-space dependent interactions are easy to include in the AFDMC propagation. The spin/isospin-dependent components of Eq.~(\ref{eq:vlij}) can be recast as a sum of quadratic operators, involving only spin and isospin of two particles at a time. This allows us to fully include the potentials described above in the imaginary time propagation of the trial state of Eq.~(\ref{eq:wf}). The sampling of spatial coordinates and auxiliary fields for the spinor rotations is done as in Ref.~\cite{Carlson:2015} via a ``plus-minus'' sampling. This allows one to sensibly lower the variance compared to the simpler sampling scheme employed in Ref.~\cite{Lonardoni:2014}. The Fermion sign problem~\cite{Carlson:2015} is controlled by constraining the evolved configurations to have positive real overlap with the trial function (constrained-path approximation)~\cite{Zhang:2003}. Expectation values of the quantities of interest are evaluated as described in Refs.~\cite{Lonardoni:2014,Carlson:2015}.

\section{Results and discussion}
\setlength{\tabcolsep}{19.5pt}
\begin{table}[b]
\centering
\caption[]{Binding energies (in MeV) for selected nuclei with $4\le (A-1)\le 48$. The experimental total angular momentum, parity, and isospin $\left(J^\pi,T\right)$ are also shown. Results are calculated by employing the two-nucleon AV4' potential plus the central component of the three-nucleon UIX force, see text for details. Experimental energies are shown for comparison.}
\begin{tabular}{lcc}
\hline
\isotope[A-1]{Z} & AV4'+UIX$_c$ & Exp \\
\hline
\isotope[2]{H}\,$\left(1^+,0\right)$                      & $-2.25(2)$  & $-2.22$  \\
\isotope[3]{H}\,$\left(\frac{1}{2}^+,\frac{1}{2}\right)$  & $-7.76(5)$  & $-8.48$  \\
\isotope[4]{He}\,$\left(0^+,0\right)$                     & $-26.63(2)$ & $-28.30$ \\
\isotope[15]{O}\,$\left(\frac{1}{2}^-,\frac{1}{2}\right)$ & $-99.6(1)$  & $-112.0$ \\
\isotope[16]{O}\,$\left(0^+,0\right)$                     & $-119.9(2)$ & $-127.6$ \\
\isotope[39]{K}\,$\left(\frac{3}{2}^+,\frac{1}{2}\right)$ & $-360.8(2)$ & $-333.7$ \\
\isotope[40]{Ca}\,$\left(0^+,0\right)$                    & $-383.3(3)$ & $-342.1$ \\
\isotope[44]{Ca}\,$\left(0^+,2\right)$                    & $-397.8(5)$ & $-381.0$ \\
\isotope[47]{K}\,$\left(\frac{1}{2}^+,\frac{9}{2}\right)$ & $-386.3(2)$ & $-400.2$ \\
\isotope[48]{Ca}\,$\left(0^+,4\right)$                    & $-413.2(3)$ & $-416.0$ \\
\hline
\end{tabular}
\label{tab:en}
\end{table}

The AFDMC binding energy for selected nuclei are reported in Tab.~\ref{tab:en}. Although the employed two- and three-body nucleon interaction model is quite simple (AV4' + UIX$_c$), AFDMC predicts binding energies of nuclei for $(A-1)<50$ within $\approx10\%$ of the experimental values. This agreement can be further improved by employing more realistic two- and three-body forces, and at the same time improving the wave functions and propagation techniques, as done, for instance, in Refs.~\cite{Lonardoni:2018prl,Lonardoni:2018prc,Lonardoni:2018rhok}. However, this is not relevant for the current work where the observable of interest is an energy difference $(B_\Lambda)$, not sensitive to the details of the employed nuclear force~\cite{Lonardoni:2013}. It is however interesting to note that the employed simplified nuclear potential retains the basic structure of a pion-less effective interaction at leading order, and it is capable of providing a reasonable description of the ground-state energy of nuclei up to calcium.

\begin{figure}[t]
	\centering
	\includegraphics[width=\linewidth]{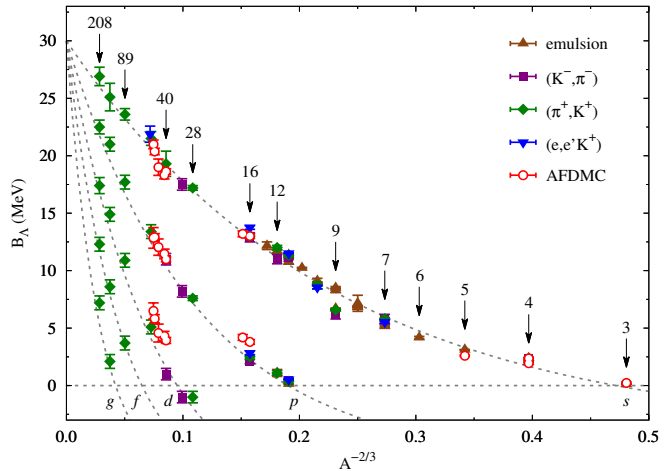}
	\caption[]{$\Lambda$~separation energy for different hyperon orbits. Solid symbols are the updated experimental results from different production mechanisms~\cite{Gal:2016}. Empty red circles are the results of this work. Dashed lines are just a guide to the eye.}
	\label{fig:bl}
\end{figure}

In Fig.~\ref{fig:bl} we present the $\Lambda$~separation energy of hypernuclei from $A=3$ to $A=49$ calculated with AFDMC for different hyperon orbits. Theoretical results are consistent with available experimental data over the entire mass range analyzed. Remarkably, even though the three-body $\Lambda NN$ force was fit to the $\Lambda$~separation energy in the $s$-orbit only, the agreement between theory and experiment for $B_\Lambda$ in different hyperon orbits is also very good. 

\setlength{\tabcolsep}{5pt}
\begin{table}[b]
\centering
\caption[]{$\Lambda$~separation energy (in MeV) for $A\le16$ in $s$- and $p$- orbits. For each hyperon orbit, AFDMC results (first column) are compared to available experimental data~\cite{Gal:2016} (second column). The quoted experimental value for \isotope[17][\Lambda]{O} in $s$-orbit is the semiempirical result of Ref.~\cite{Usmani:1995}. For each system, the total angular momentum, parity, and isospin $\left(J^\pi,T\right)$ of the dominant peaks (measured or expected) are also shown.}
\begin{tabular}{lcccc}
\hline
\isotope[A][\Lambda]{Z} & \multicolumn{2}{c}{$s$-orbit} & \multicolumn{2}{c}{$p$-orbit} \\
\hline
\isotope[3][\Lambda]{H}  & $\left(\frac{1}{2}^+,\frac{1}{2}\right)$ $0.23(9)$ & $0.13(5)$ & & \\
\isotope[4][\Lambda]{H}  & $\left(0^+,\frac{1}{2}\right)$ $1.95(9)$ & $2.16(8)$ & & \\
\isotope[4][\Lambda]{He} & $\left(0^+,\frac{1}{2}\right)$ $2.37(9)$ & $2.39(3)$ & & \\
\isotope[5][\Lambda]{He} & $\left(\frac{1}{2}^+,0\right)$ $2.75(5)$ & $3.12(2)$ & & \\
\isotope[16][\Lambda]{O} & $\left(1^-,\frac{1}{2}\right)$ $13.0(2)$ & $13.0(2)$ & $\left(2^+,\frac{1}{2}\right)$ $3.8(2)$ & $2.5(2)$ \\
\isotope[17][\Lambda]{O} & $\left(\frac{1}{2}^+,0\right)$ $13.2(3)$ & $13.0(4)^*$ & $\left(\frac{3}{2}^-,0\right)$ $4.2(3)$ & --- \\
\hline
\end{tabular}
\label{tab:bl1}
\end{table}

\setlength{\tabcolsep}{4.5pt}
\begin{table}[t]
\centering
\caption[]{$\Lambda$~separation energy (in MeV) for $A\ge40$. AFDMC results at the top, experimental data at the bottom. Data for \isotope[40][\Lambda]{Ca} and \isotope[51][\Lambda]{V} are taken form Ref.~\cite{Gal:2016}. The $s$-orbit $B_\Lambda$ for \isotope[40][\Lambda]{Ca} and \isotope[52][\Lambda]{V} are from Ref.~\cite{Pile:1991} and Ref.~\cite{Gogami:2014}, respectively. For each system, the total angular momentum, parity, and isospin $\left(J^\pi,T\right)$ of the dominant peaks (measured or expected) are also shown.}
\begin{tabular}{lccc}
\hline
\isotope[A][\Lambda]{Z} & $s$-orbit & $p$-orbit & $d$-orbit \\
\hline
\isotope[40][\Lambda]{K}  & $\left(2^+,\frac{1}{2}\right)$ $18.6(3)$ & $\left(3^-,\frac{1}{2}\right)$ $11.0(3)$ & $\left(4^+,\frac{1}{2}\right)$ $3.9(3)$ \\
\isotope[41][\Lambda]{Ca} & $\left(\frac{1}{2}^+,0\right)$ $18.3(4)$ & $\left(\frac{3}{2}^-,0\right)$ $11.5(4)$ & $\left(\frac{5}{2}^+,0\right)$ $4.3(4)$ \\
\isotope[45][\Lambda]{Ca} & $\left(\frac{1}{2}^+,2\right)$ $19.0(8)$ & $\left(\frac{3}{2}^-,2\right)$ $12.1(8)$ & $\left(\frac{5}{2}^+,2\right)$ $4.6(8)$ \\
\isotope[48][\Lambda]{K}  & $\left(1^+,\frac{9}{2}\right)$ $20.4(4)$ & $\left(2^-,\frac{9}{2}\right)$ $12.9(4)$ & $\left(3^+,\frac{9}{2}\right)$ $5.8(4)$ \\
\isotope[49][\Lambda]{Ca} & $\left(\frac{1}{2}^+,4\right)$ $21.0(5)$ & $\left(\frac{3}{2}^-,4\right)$ $12.9(9)$ & $\left(\frac{5}{2}^+,4\right)$ $6.5(7)$ \\
\noalign{\smallskip}\hline\noalign{\smallskip}
\isotope[40][\Lambda]{Ca} & $\left(2^+,\frac{1}{2}\right)$ $19.3(1.1)$ & $\left(3^-,\frac{1}{2}\right)$ $11.0(5)$ & $\left(4^+,\frac{1}{2}\right)$ $1.0(5)$ \\
\isotope[51][\Lambda]{V}  & $\left(?,2\right)$ $21.5(6)$               & $\left(?,2\right)$ $13.4(6)$             & $\left(?,2\right)$ $5.1(6)$ \\
\multirow{2}{*}{\isotope[52][\Lambda]{V}} & $\left(3^-,\frac{5}{2}\right)$ $21.9(7)$ & --- & --- \\
                                          & $\left(4^-,\frac{5}{2}\right)$ $21.9(7)$ & --- & --- \\
\hline
\end{tabular}
\label{tab:bl2}
\end{table}

We report in Tabs.~\ref{tab:bl1} and \ref{tab:bl2} details of the AFDMC calculations for $\Lambda$~hypernuclei. In the former, the $\Lambda$~separation energy for $s$- and $p$-shell hypernuclei is shown. In the latter, the $B_\Lambda$ values of the expected dominant peaks in $s$-, $p$-, and $d$-orbits are reported for hypernuclei with $40\le A\le49$, in comparison with the limited available experimental information for the neighboring hypernuclei \isotope[40][\Lambda]{Ca}, \isotope[51][\Lambda]{V}, and \isotope[52][\Lambda]{V}. In the light sector, AFDMC results are in general compatible with observations. The $\Lambda$~separation energy of \isotope[4][\Lambda]{H} is slightly lower than the current experimental value. However, it is consistent with the old emulsion data of $2.04(4)\,\rm MeV$, to which the employed spin-independent CSB potential was originally fit~\cite{Usmani:1999}. In the medium-mass region for $A\ge40$, results for different hyperon orbits show a consistent pattern, and they are compatible with the experimental data for the nearest hypernucleus. Exception is $B_\Lambda$ in the $d$-orbit for $A\sim40$, for which the only available experimental value indicates a much smaller $\Lambda$~separation energy. However, this result is from the $(K^-,\pi^-)$ exchange reaction on \isotope[40]{Ca} performed at Saclay in 1979~\cite{Bertini:1979} and never confirmed in following experiments. Theoretical calculations of the photoproduction cross section of \isotope[40][\Lambda]{K}~\cite{Bydzovsky:2012} indicate a value for $B_\Lambda$ in the $d$-orbit much closer to the one of this work. This, in addition to the very low statistics for the $(\pi^+,K^+)$ production of \isotope[40][\Lambda]{Ca}~\cite{Pile:1991}, emphasizes the need of accurate experimental research in the medium-mass region of the hypernuclear chart.

In view of the hypernuclear experimental program at JLab, aimed at the high precision spectroscopy of $\Lambda$ hypernuclei for $A=40,48$ with an electron beam, the AFDMC predictions for \isotope[40][\Lambda]{K} and \isotope[48][\Lambda]{K} are of particular interest.  Statistical uncertainties due to the use of a Monte Carlo procedure for $A=40,48$ are smaller than those of current nearby experimental data. Quantum Monte Carlo could then be used to extract accurate information on the hyperon-nucleon force, if supported by more accurate measurements of the $\Lambda$~separation energy. In this particular regime, it is possible to start addressing the fundamental problem of the nucleon-isospin dependence of the three-body $\Lambda NN$ interaction, which is of great importance in extrapolating the experimental information on hypernuclei to the study of NS matter. 

\begin{figure}[b]
	\centering
	\includegraphics[width=0.8\linewidth]{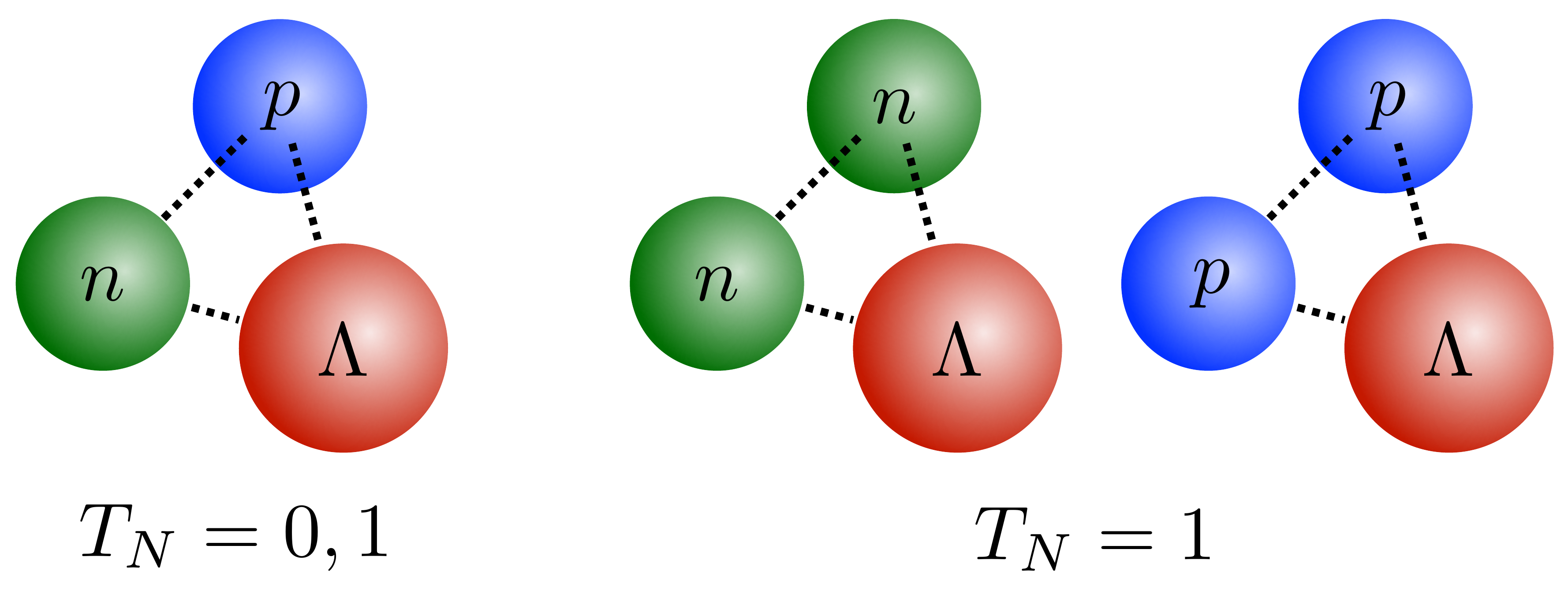}
	\caption[]{Nucleon-isospin $T_N=0$ and $T_N=1$ configurations in $\Lambda NN$ triplets.}
	\label{fig:lnn_trip}
\end{figure}

The current version of the $\Lambda NN$ potential does not depend on whether the two nucleons are in a isospin singlet or a isospin triplet state  (see Fig.~\ref{fig:lnn_trip}). For symmetric hypernuclei the Pauli principle suppresses any strong contribution from the $\Lambda nn$ or $\Lambda pp$ channels. On the other hand, in hyper-neutron matter or in matter at $\beta$-equilibrium, the contribution of the nucleon-isospin triplet channel might become relevant. By looking at the nucleon-isospin asymmetry, defined as $\delta=(N-Z)/A$ with $N$ the number of neutrons and $Z$ that of protons, \isotope[40][\Lambda]{K} and \isotope[48][\Lambda]{K} are characterized by having the smallest and largest nucleon-isospin asymmetry along the hypernuclear chart. Moreover, they are large enough to manifest central densities comparable to saturation density, but they are still accessible using ab-initio techniques such as AFDMC. $A=40,48$ $\Lambda$~hypernuclei are thus excellent candidates to be used to extract information on the hypernuclear force relevant for the description of strange neutron-rich nuclei and for the prediction of NS properties.

In order to test the sensitivity of our AFDMC predictions to a possible nucleon-isospin dependence of the hypernuclear force, we performed a proof-of-principle study. The first two terms of Eq.~(\ref{eq:vlij}) depend upon the nucleon-isospin via the operator $\bm\tau_i\cdot\bm\tau_j$. This can be written in terms of the projectors on the singlet and triplet nucleon-isospin channels, $\bm\tau_i\cdot\bm\tau_j=-3\,\mathcal P_{ij}^{T_N=0}+\mathcal P_{ij}^{T_N=1}$. The nucleon-isospin dependent terms of Eq.~(\ref{eq:vlij}) can then be expressed as:
\begin{align}
	v_{\lambda ij}^T\,\bm\tau_i\cdot\bm\tau_j=-3\,v_{\lambda ij}^T\,\mathcal P_{ij}^{T_N=0}+(1+C_T)\,v_{\lambda ij}^{T}\,\mathcal P_{ij}^{T_N=1},
    \label{eq:proj}
\end{align}
where the additional parameter $C_T$ has been introduced in order to single out the nucleon-isospin triplet contribution compared to the original formulation of Eq.~(\ref{eq:vlij}). 

By (artificially) varying the $C_T$ parameter we can test whether our AFDMC predictions for $B_\Lambda$ are affected by an unbalance in the strength of the nucleon-isospin triplet component compared to the nucleon-isospin singlet one. In other words, the aim of this study is to answer the following question: can AFDMC calculations be used to resolve, at least at this elementary level, the nucleon-isospin dependence of the $\Lambda NN$ force starting from experimental hypernuclear data only? Isospin-triplet components have to play a crucial role in strange neutron-rich systems, such as $A=48$ hypernuclei. What is not obvious is whether AFDMC calculations for such large systems have a chance to appreciate this effect (considering Monte Carlo statistical errors). We point out once more, that, besides the implications in hypernuclear physics, resolving this dependence has a potential strong impact on the study of neutron star matter.

\begin{figure}[t]
	\centering
	\includegraphics[width=\linewidth]{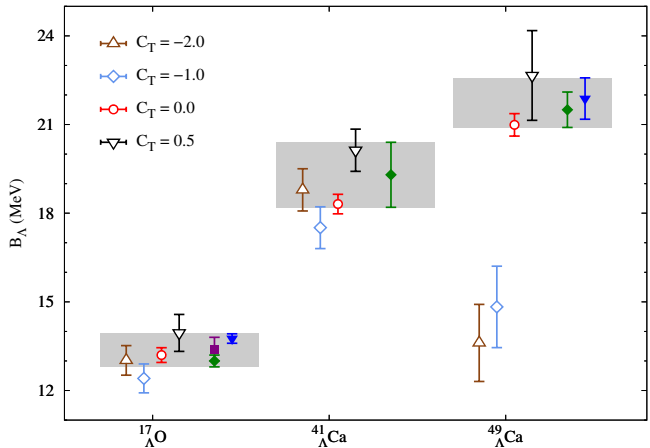}
	\caption[]{$\Lambda$~separation energy in $s$-orbit for medium-mass hypernuclei. Empty symbols are the AFDMC results for different choices of the parameter $C_T$. Solid symbols are available experimental data (same color scheme of Fig.~\ref{fig:bl}): $^{16}_{~\Lambda}$O from $(\pi^+,K^+)$~\cite{Gal:2016} and $(K^-,\pi^-)$~\cite{Agnello:2011}; $^{16}_{~\Lambda}$N from $(e,e'K^+)$~\cite{Cusanno:2009}; \isotope[40][\Lambda]{Ca} from $(\pi^+,K^+)$~\cite{Pile:1991}; \isotope[51][\Lambda]{V} from $(\pi^+,K^+)$~\cite{Gal:2016}; \isotope[52][\Lambda]{V} from $(e,e'K^+)$~\cite{Gogami:2014}. Shaded areas are current experimental constraints in the regions $A\sim16,40,48$.}
	\label{fig:bl_ct}
\end{figure}

In Fig.~\ref{fig:bl_ct} we show the results of such a sensitivity study. The $\Lambda$~separation energy for \isotope[17][\Lambda]{O}, \isotope[41][\Lambda]{Ca}, and \isotope[49][\Lambda]{Ca} is reported for different choices of the $C_T$ parameter. For $C_T=0$ the original parametrization of the three-body force is recovered, and the results are the same as in Fig.~\ref{fig:bl}. For $C_T=-1$ the nucleon-isospin triplet component of Eq.~(\ref{eq:proj}) is turned off, for $C_T=0.5$ it is enhanced, and for $C_T=-2$ it changes sign. Notice that for $C_T\neq 0$  Monte Carlo errors are larger. This is not a consequence of the introduction of the $C_T$ parameter in Eq.~(\ref{eq:proj}), rather a matter of statistics. Simulations for $C_T\neq0$ have been performed sampling a reduced number of configurations compared to those for $C_T=0$. Considering the proof-of-principle nature of this study and the size of experimental uncertainties on $B_\Lambda$, this does not affect the conclusions that one can extract from Fig.~\ref{fig:bl_ct}.

As one could have expected, for symmetric hypernuclei the variation of $B_\Lambda$ due to the change of $C_T$ is not large, in particular for the lightest system. All results obtained for different values of $C_T$  fall within the uncertainty band determined by the available experimental data for neighboring hypernuclei, despite the large variation of the control parameter $(C_T\in[-2.0,0.5])$. The case of \isotope[49][\Lambda]{Ca} differs; one finds a large nucleon-isospin asymmetry $\delta\approx0.16$, for which the variation of the control parameter greatly affects the prediction of $B_\Lambda$, in particular for $C_T<0$. Even though the range of $C_T$ considered here is unrealistically large, the results for $A=49$ suggest that AFDMC calculations for the employed hypernuclear potentials are sensitive to a possible nucleon-isospin dependence of the three-body hyperon-nucleon force for $A\gtrsim40$ and $\delta\neq0$.

However, the current experimental information in the region of interest $(40\lesssim A\lesssim50)$ is not sufficiently accurate, as clearly visible from the gray bands in Fig.~\ref{fig:bl_ct}. The high precision spectroscopy of \isotope[40][\Lambda]{K} and \isotope[48][\Lambda]{K} that will be performed in the coming years at the Thomas Jefferson Laboratory will provide a new and valuable set of constraints to be used in the construction of realistic hyperon-nucleon interactions. The accuracy of the expected experimental data should be sufficient to allow one to re-fit the $W_D$ and $C_T$ parameters of Eqs.~(\ref{eq:vlij}) and (\ref{eq:proj}) in order to simultaneously reproduce the $\Lambda$~separation energy of \isotope[40][\Lambda]{K} and \isotope[48][\Lambda]{K} with AFDMC calculations. This will provide a unique opportunity to investigate how the nucleon-isospin dependence of a many-body hyperon-nucleon force affects the hypernuclear spectrum, in particular for neutron-rich systems. It could elucidate the role of three-body forces in the determination of the unusually large spin-dependent hypernuclear CSB in light nuclei~\cite{Esser:2015,Yamamoto:2015,Botta:2017}, and it could shed some light on the implications for the EoS of hypernuclear matter and for the prediction of NS properties.

\section{Conclusions}
We have performed quantum Monte Carlo calculations of single-$\Lambda$~hypernuclei in the mass range $3\le A\le 49$ using phenomenological two- and three-body hypernuclear interactions. The accuracy of these calculations allows for the prediction of $\Lambda$~separation energies with statistical errors comparable or even smaller than current experimental uncertainties, especially for medium-mass hypernuclei. From a sensitivity study we can learn that quantum Monte Carlo methods can be used to resolve a possible nucleon-isospin dependence of the three-body $\Lambda NN$ force starting from hypernuclear data. We conclude that future hypernuclear experiments on medium-mass targets will provide an opportunity to develop more accurate many-body hyperon-nucleon interactions, crucial for the description of the physics of hypernuclei and even more for the prediction of neutron star properties, for which new and more strict constraints have become available~\cite{Demorest:2010,Antoniadis:2013,Fattoyev:2018}.\\

\noindent Preliminary results for the nucleon-isospin dependence of three-body hyperon-nucleon-nucleon force were obtained by F.~Catalano during the initial phase of this work. We thank S.~Reddy, S.~Gandolfi, A.~Lovato, L.~Contessi, A.~Gal, S.~N.~Nakamura, and H.~Tamura for many valuable discussions. The work of D.L. was supported by the U.S. Department of Energy, Office of Science, Office of Nuclear Physics, under Contract No. DE-SC0013617, and by the NUCLEI SciDAC program. Computational resources have been provided by the National Energy Research Scientific Computing Center (NERSC), which is supported by the U.S. Department of Energy, Office of Science, under contract DE-AC02-05CH11231, and by Los Alamos Open Supercomputing via the Institutional Computing (IC) program.

\end{document}